# Total ionizing dose effects of domestic SiGe HBTs under different dose rate

Liu Mo-Han(刘默寒)[1,2] Lu Wu(陆妩)[1,2†] Ma Wu-Ying(马武英)[1] Wang Xin(王信)[1] Guo Qi(郭旗)[1]

He Cheng-Fa(何承发)[1] Jiang Ke(姜柯)[1] Li Xiao-Long(李小龙)[1] Xiong Ming-Zhu(荀明珠)[1,2]

[1]Key Laboratory of Functional Materials and Devices for Special Environments, Xinjiang Key Laboratory of Electronic Information Materials and Devices, Xinjiang Technical Institute of Physics and Chemistry, Chinese Academy of Sciences, Urumqi 830011, China

[2]School of Physics Science and Technology, Xinjiang University, Urumqi 830046, China

**Abstract:** The total ionizing radiation (TID) response of commercial NPN silicon germanium hetero-junction bipolar transistors (SiGe HBTs) produced domestic were investigated under the dose rate of 800mGy(Si)/s and 1.3mGy(Si)/s with Co-60 gamma irradiation source, respectively. The changes of the transistor parameter such as Gummel characteristics, excess base current before and after irradiation are examined. The results of the experiments show that for the KT1151, the radiation damage has been slightly different under the different dose rate after the prolonged annealing, shows a time dependent effect(TDE). But for the KT9041, the degradations of low dose rate irradiation is higher than the high dose rate, demonstrate that there have a potential enhanced low dose rate sensitive(ELDRS) effect exist on KT9041. The possible underlying physical mechanisms of the different dose rates response induced by the gamma ray are discussed.

## 1 Introduction

In the last twenty years, Silicon germanium (SiGe) hetero-junction bipolar transistors (HBTs) technology has been considered to be one of the promising candidate for future space applications due to its exciting high current gain, low noise response and cryogenics temperature performance, especially the built-in tolerance of total ionizing dose (TID) radiation and displacement damage (DD)[1]. A lot of literatures have been published about the radiation tolerance of SiGe HBTs with various radiation source, bias, device structures and dose rate[2-7]. However, most of the experiments performed with the dose rate above 500mGy(Si)/s which is greatly higher than that of actual space radiation environment as low as $10^{-3}$mGy(Si)/s~0.1mGy(Si)/s, induced a risk of overestimate the radiation resistance of SiGe HBTs with the high dose rate to evaluate the radiation-hard performance of SiGe HBTs. Therefore, it is necessary to carry out the experiments in low dose rate condition that can approximately simulate the actual space low dose rate environment.

In this paper, we performed the radiation experiments of two domestic commercial NPN SiGe HBTs devices with gamma ray under the two different dose rate, detailed investigated the dose rate response of the domestic commercial SiGe HBTs. The measurements result of the direct current (DC) parameters shows that different device have different performance degradation under the different dose rate irradiation, the KT9041 shows more resistant to radiation compared to KT1151, but the KT9041 experienced more serious degradation under the low dose rate irradiation than the degradation of high dose rate, shows there may have an enhanced low dose rate sensitive (ELDRS) effect in KT9041. Then, the potential physical mechanisms of the different dose rates response induced by the gamma ray are detailed discussed.

## 2 Experiments

The devices investigated in this work were two kinds of NPN silicon germanium hetero-junction bipolar transistors (SiGe HBTs) designed and fabricated domestic, which are designed for high frequency low noise amplifier with the advantages of low noise figure, high power gain, high voltage, broad dynamic range and good linearity. The major characteristics of the two devices are listed in the Table 1.

The irradiations experiments described in this paper are performed at room temperature with 60-Co water-well gamma irradiation source in Xinjiang Technical Institute of Physics & Chemistry of Chinese, Academy of Sciences. The high dose rate(HDR) and low dose rate(LDR) used in the experiments are 800mGy (Si)/s and 1.3mGy(Si)/s, respectively. The devices were mounted in the irradiation boards with all terminals grounded ($V_B$=$V_E$=$V_C$=0.0V) during the irradiation and annealing process, and irradiated to a maximum total ionizing dose level of 11kGy(Si). The electrical parameters including Gummel characteristics and direct current gain of the devices utilized to characterize the degradation were measured with KEITHLEY 4200-SCS Semiconductor Parameter Analyzer removed from the irradiation room within 20 minutes at room temperature before and after each specified value of accumulated dose.

Table 1. Characteristics of the devices

| SAMPLE | Polarity | Ic(mA) | Vceo(V) | $f_T$(GHz) | β |
|---|---|---|---|---|---|
| KT9041 | NPN | 30 | 4.5 | 25 | 150 |
| KT1151 | NPN | 20 | 12 | 7 | 300 |

## 3 Results and Discussion

### 3.1 Degradation of Base Current and Current Gain

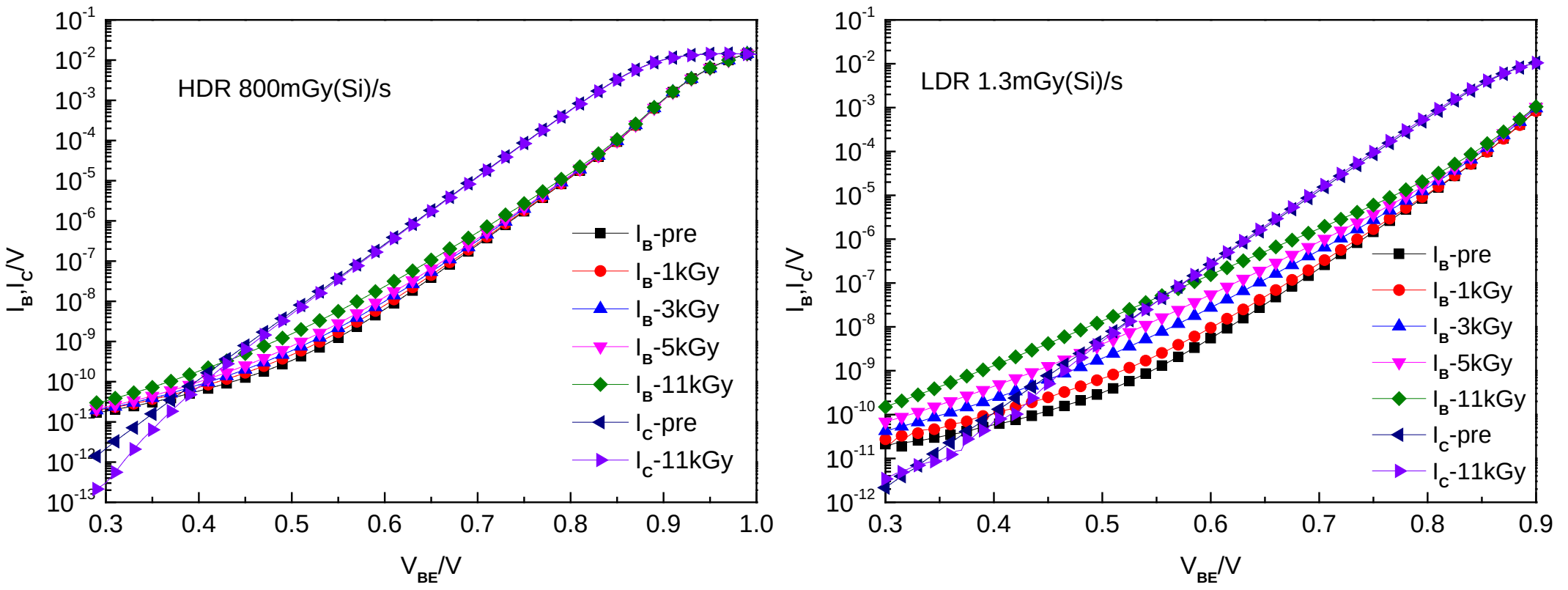

Fig.1. Forward-mode Gummel characteristics of KT9041 as a function of total dose

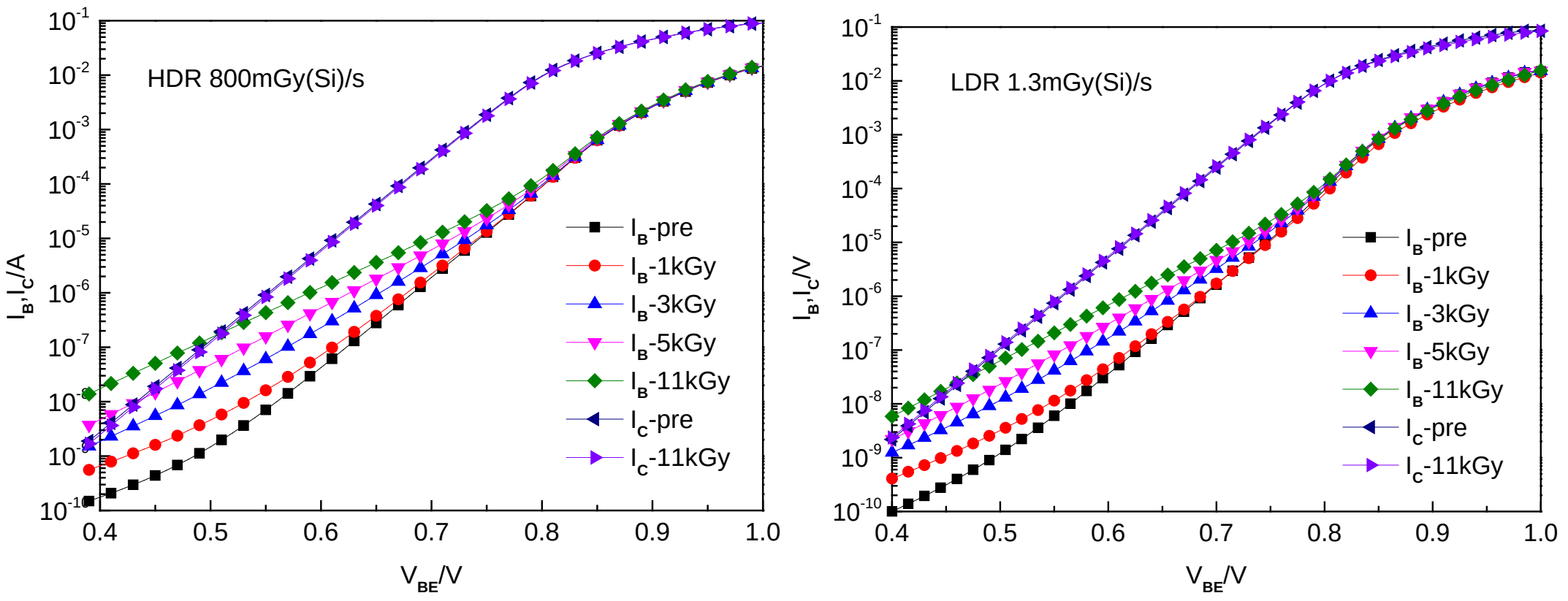

Fig.2. Forward-mode Gummel characteristics of KT1151 as a function of total dose

Figure 1 and figure 2 presents the Gummel characteristics of the two devices as a function of accumulated total dose under two different dose rates. For the two devices, the base currents $I_B$ are both monotonously increasing with the increased accumulated total dose under the two different high and low dose rate, especially at the low base-emitter junction voltage($V_{BE}$＜0.8V). While compared with IB, the collector current IC remains the same, only have slight changes. Thus, direct current gains($\beta= I_C/I_B$) of the two devices are decreased after the irradiation under the two high and low dose rates. These experiments phenomena indicate that the base current $I_B$ is more sensitive to the radiation damage caused by Co-60 gamma irradiation, while the collector current $I_C$ is only slightly affected by irradiation at the given base-emitter voltage value ($V_{BE}$). Also can be seen clearly from the figure 1 and figure 2 is that the changes of the $I_B$ of the KT1151 are greater than that of KT9041 under the different dose rate. And compared the degradation of $I_B$ under the high and low dose rates, the changes of the $I_B$ under the high dose rate are small than the changes under the low dose rate.

In order to quantitatively compare the effect of the degradation of the devices induced by the radiation under different dose rate, we defined two parameters，the excess base current $\Delta I_B$( $\Delta I_B=I_{B\text{-post}}-I_{B\text{-pre}}$) and Normalized current gain $\beta_{post}/\beta_{pre}$， in which the $I_{B\text{-pre}}$ , $I_{B\text{-post}}$ and $\beta_{pre}$， $\beta_{post}$ are corresponding to the base current $I_B$ and direct current gain $\beta$ extracted from the figure 1 and figure 2 at the $V_{BE}$=0.7 before and after the irradiation.

Figure 3 and Figure 4 shows the changes of the excess base current $\Delta I_B$ of KT9041 and KT1151 as a function of total dose and annealing time under the different dose rate, respectively. With the dose accumulating, the $\Delta I_B$ increasing under all two dose rates at the voltage of 0.7V. However, there have a great deal of difference between the KT9041 and KT1151. For the KT9041，It can be seen clearly from Figure 3 that the changes of $\Delta I_B$ are approximately two orders of magnitude after the dose up to 11kGy(Si) under the low dose rate of 1.3mGy(Si)/s, while the changes of $\Delta I_B$ under high dose rate of 800mGy(Si)/s are much less than that of low dose rate and there have some extent of increasing trend with the annealing time after the irradiation which indicate that there may have 'post-radiation damage' exist in the KT9041. Contrary to the KT9041，the $\Delta I_B$ of KT1151 under the high dose rate is greater than the $\Delta I_B$ under the low dose rate. And the $\Delta I_B$ decreased dramatically at the first few decade hours of annealing，and then decreasing slowly down to the value that smaller than the value of $\Delta I_B$ irradiated to 11kGy(Si) under the low dose rate with the increasing annealing time.

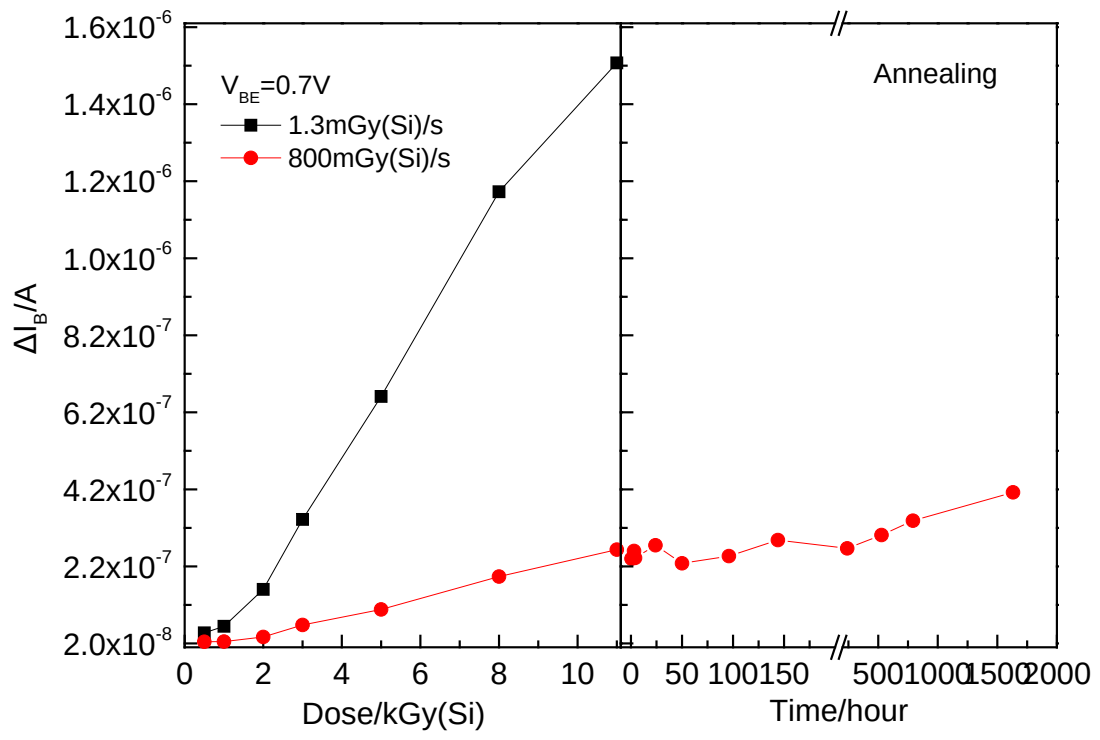

Fig.3. Excess base current of KT9041 as a function of total dose and annealing time

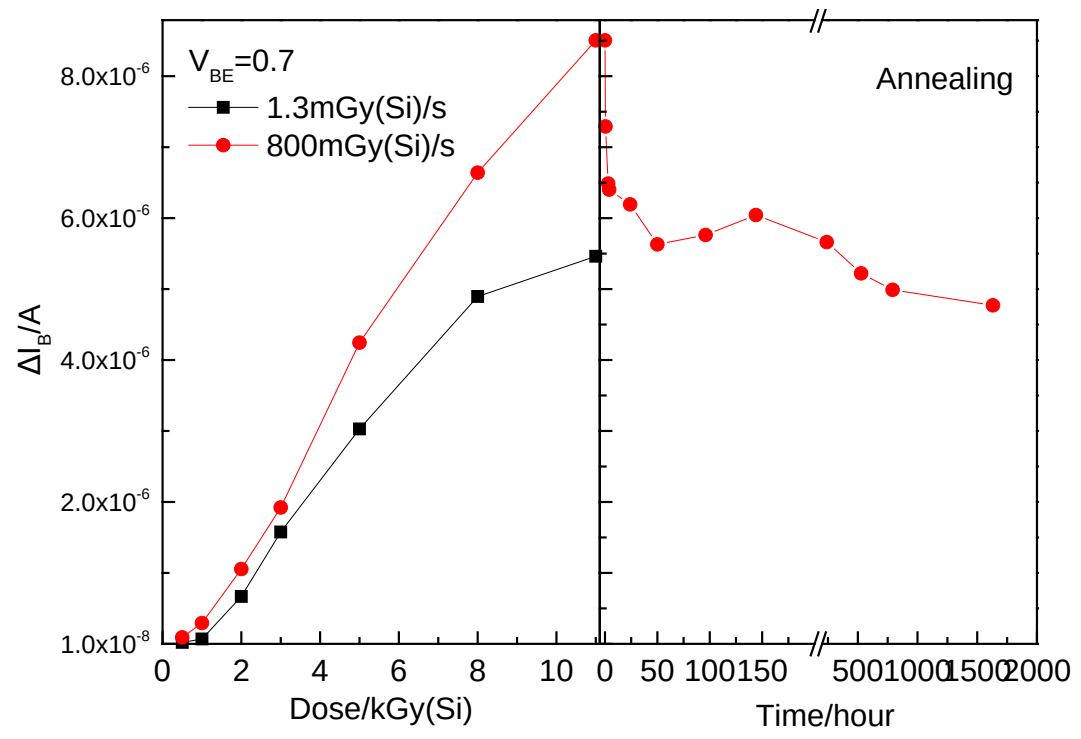

Fig.4. Excess base current of KT1151 as a function of total dose and annealing time

The normalized current gain of KT9041 and KT1151 as a function of total dose and annealing time are shown in Figure 5 and Figure 6 under the different dose rate. The normalized current gains of the two devices are monotonically decreasing with the increasing irradiation dose under the different dose rate. The difference is that the degradation of normalized current gain of KT9041 under low dose rate is greater than the degradation under high dose rate，while to the KT1151，the degradation of normalized current gain almost the same after irradiation to the dose of 11kGy(Si). The degradation of normalized current gain under the high dose rate is nearly stayed the same as the value irradiation to the 11kGy(Si) with the annealing time increasing for the KT9041 which shows a clearly effect of enhanced low dose rate sensitive(ELDRS). For the KT1151，It recovering quickly to the value irradiation to the 11kGy(Si) under the low dose rate，and then remains unchanged which shows there have a time dependent effect(TDE) exists in the KT1151.

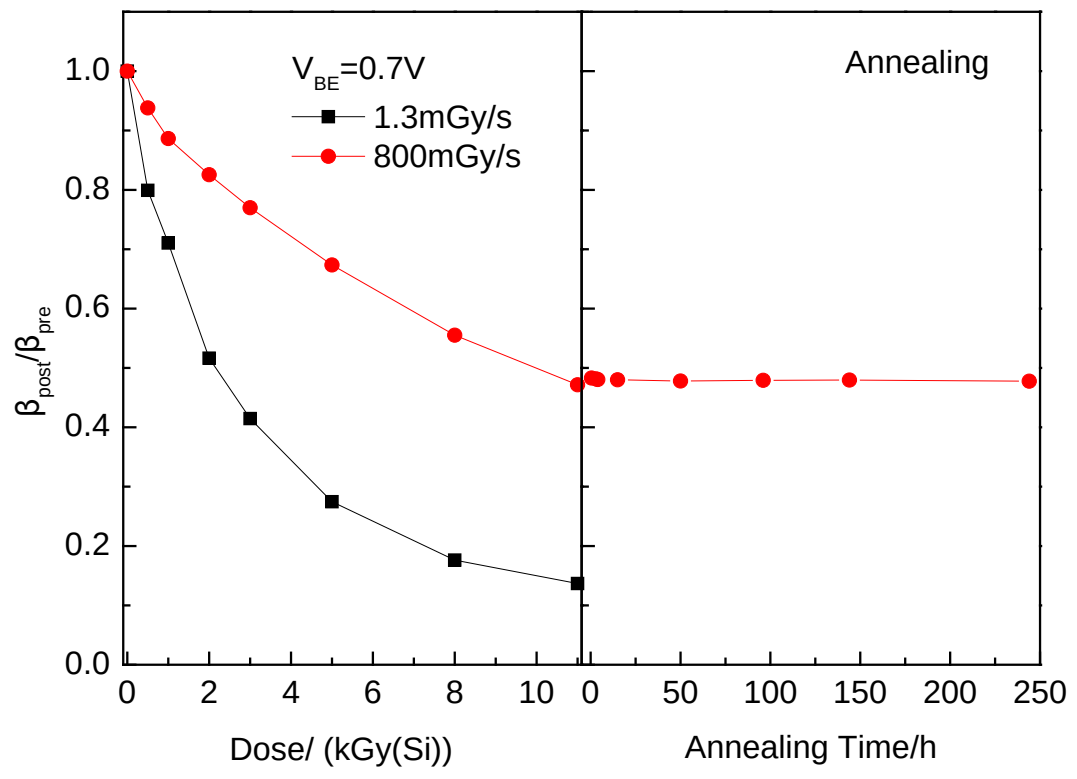

Fig.5. Normalized current gain of KT9041 as a function of total dose and annealing time

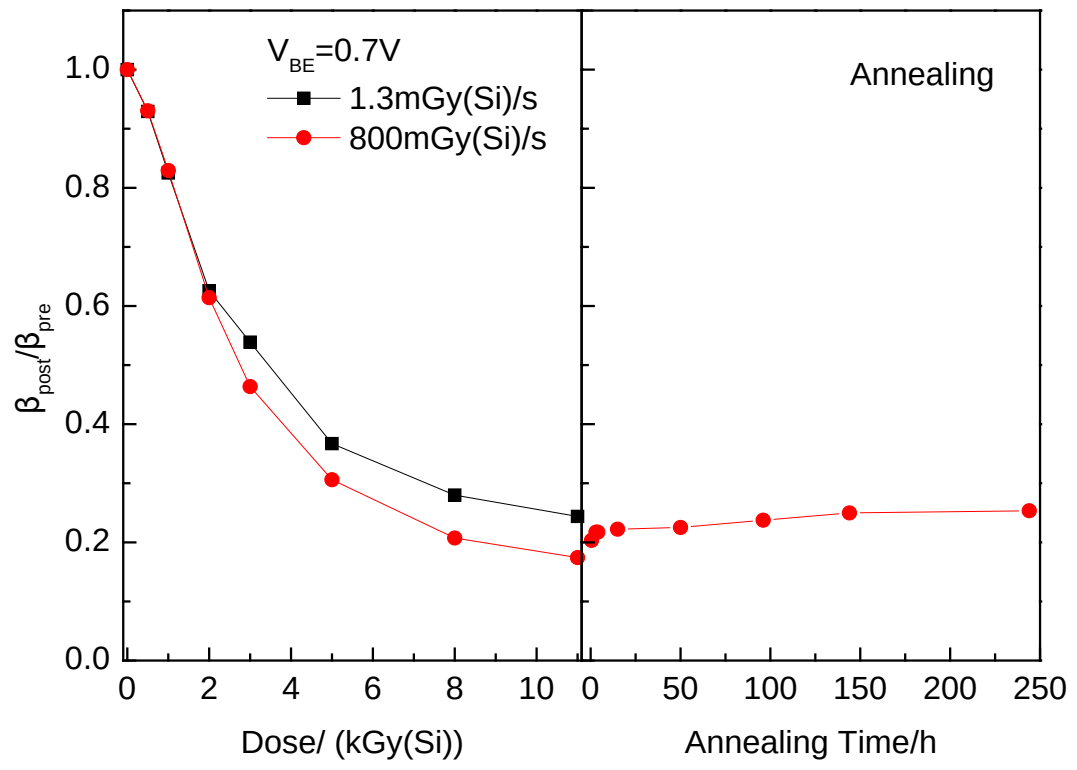

Fig.6. Normalized current gain of KT1151 as a function of total dose and annealing time

## 3.2 Discussion

### 3.2.1 Total ionizing effects

The above experimental results show that the base current increases with increasing accumulated dose for the high and low dose rates irradiation, causing a significant drop in current gain both for the KT9041 and KT1151. Generally speaking，the degradation of the base current of the SiGe HBTs are due to increase of the surface recombination current( $I_{bsr}$ ) in base region which is related to the radiation induced oxide trap charge( $N_{ot}$ ) and interface trap charge( $N_{it}$ ) in the SiGe HBTs base-emitter spacer oxide and interface of Si /$SiO_2$[2]. The surface recombination current can be approximately characteristics by the equation below[8]，

$$I_{bsr} \sim N_{it} \exp(\alpha \mathrm{N}_{ot}^2). \qquad (1)$$

where the $\alpha = 1/2q\varepsilon\varepsilon_0 N_a$ is connected with the electronic charge $q$, absolute dielectric constant $\varepsilon_0$, relative dielectric constant $\varepsilon$ and the doping of the substrate $N_a$. Thus, the more the radiation induced oxide trap and interface trap, the more the degradation of surface recombination current $I_{bsr}$.

The effect of the radiation induced surface trap charge to the base current can be expressed by the surface recombination velocity( SRV )[9]. In the low injection condition, the relationship between surface recombination velocity SRV and surface trap charge $N_{it}$ approximated as

$$\mathrm{SRV} \cong \sigma N_{it} V_{th}. \qquad (2)$$

Where $\sigma$ is the trap capture cross section and $V_{th}$ is the thermal velocity of carriers in silicon. The radiation induced buildup in $N_{it}$ increases base current in the SiGe HBTs by increasing the surface recombination rate. The radiation induced excess base current $\Delta I_B$ can be expressed as a function of SRV by the following equation,

$$\Delta I_B = q\int_s \Delta U ds = q\Delta SRV \int_s \frac{(np - n_i^2)}{(n + p + 2n_i)} ds. \qquad (3)$$

where $q$ is the electronic charge, $\Delta U$ is the change of the surface recombination rate induced by radiation, $s$ is the recombination surface area, and $n$、$p$、$n_i$ is the concentration of the electron、hole and intrinsic carrier, respectively. Therefore, the accumulation of $N_{it}$ not only can cause a linear increase in surface recombination velocity SRV according to the equation (2), and also can affect the carrier concentration of the device according to the equation (3) induced a non-linear increase of the $\Delta I_B$ as shown in Figure 3 and Figure 4. And finally, cause the drops of the current gains of the two devices under different dose rate.

Both of the oxide trap charge and interface trap charge can contribute to the degradation of the devices, but there still have some differences during the post-irradiation annealing. Due to the competition between the oxide trap charge and interface trap charge according to the literature[10], there has more interface trap charge induced by the prolonged low dose rate irradiation, while more oxide trap charge induced by the high dose rate. And for the different devices fabricated by the different process, the amounts of the oxide trap charge and interface trap charge induced by radiation may different because of the amounts of the defects introduced in the oxide. The oxide trap charge can be eliminated by annealing at room temperature, but the interface trap charge cannot be significantly removed by annealing under the 100ºC, thus caused a different annealing performance of the two devices under the different dose rate[11], as shown in Figure 3—Figure6.

### 3.2.2 Dose rate effects

The dose rate effects of the two different devices are investigated. As shown in the Figure 3 and Figure 5 for the KT9041, the lower the dose rate, the higher the radiation damage, and after the prolonged annealing there have only slightly recover for the $\Delta I_B$, which demonstrates a significantly enhanced low dose rate sensitivity (ELDRS) effect. As for the KT1151 in Figure 4 and Figure 6, just contrary to the KT9041, with the increasing annealing time the difference between the high dose rate irradiation and low dose rate is narrowing down and finally approximately stay the same as the degradation under the low dose rate, which shows a time dependent effect(TDE).

According to the ELDRS theory of traditional Si BJT (Bipolar Junction Transistor), compared with the low dose rate irradiation, the high dose rate irradiation induced excess holes can react with defects and releasing neutral hydrogen atoms[12-14]. The released hydrogen atoms combine into molecular hydrogen, and then interact with the neutral trap site and in the process, protons are released. The produced protons can combine with the interface defect near the Si/SiO2 interface into interfacial state. But, the hydrogen can fast diffuse to the interface of the Si/SiO2 and passivate the dangling bonds due to the lower diffusion barrier of the molecular hydrogen in SiO2, reduce the concentration of the interfacial state, thus suppress the further increasing of the base current.

Contrary to the high dose rate irradiation, when the irradiation is carried out under low dose rates, the combination between the radiation induced hydrogen is difficult because of the lower yield of the hydrogen atoms under the low dose rate. Most of them are capture holes and release protons, and then coupled with the dangling bonds formed a great deal of interfacial states. Consequently, the concentration of the interfacial state of the low dose rate irradiation is higher than that of the high dose rate irradiation, thereby caused a ELDRS effects.

However, the literatures [2] and [15] considered that because of the thin emitter-base (EB) spacer oxide and heavily doped base region of the SiGe HBTs, the leakage of the base current results from the interface traps that induced by the irradiation is suppressed, which results in an excellent performance of radiation-hardness of TID. And, strictly controlled process and special steps make sure that every hydrogen contaminants which are introduced in by the epitaxial Si growth are eliminated，bring about an immunity to ELDRS effects. For the two commercial SiGe HBTs devices investigated in this paper, due to the specific manufacturing process is still unclear, temporarily we can't give an explicit conclusion that the ELDRS effects is caused by the different amounts of interfacial states which related to the introduced hydrogen. Therefor, based on the discussions above, the different dose rate response between the different devices and manufacturer may relate to the interfacial traps at the $Si/SiO_2$ interface induced by the TID exposure.

## 4 Conclusion

In this paper, we investigated the degradation of electric parameters and annealing behavior of two commercial SiGe HBTs with 800mGy(Si)/s and 1.3mGy(Si)/s Co-60 gamma total ionization dose irradiation. The radiation sensitive electric parameters of the both two SiGe HBTs, like many Si BJTs, are base current and current gain. Based on the obtained experiment results, we discussed the dose rate dependence of radiation tolerance of the SiGe HBTs and concluded that there have an obvious ELDRS effects exist in the KT9041, while there have only TDE effects for the KT1151. The different radiation responses of the different SiGe HBTs under the high and low dose rate may due to the different amounts defects that introduced by the manufacturing process. And the different radiation responses between the different devices and manufacturer need further more deep researches by taking the different manufacturing process and devices structures into consideration.

## References

1 Cressler J D. IEEE Trans Microwave Theory Tech, 1998, 46(5): 572

2 Cressler J D. IEEE Trans Nucl Sci, 2013, 60(3): 1992

3 Babcock J A, Cressler J D, Vempati L S, et al. IEEE Trans Nucl Sci, 1994, 42(6): 1558

4 Zhang S, Niu G, Cressler J D, et al. IEEE Trans Nucl Sci, 2000, 47(6): 2521

5 Cressler J D, Krithivasan R, Sutton A K, et al. IEEE Trans Nucl Sci, 2003, 50(6): 1805

6 Haugeruda B M, Pratapgarhwalaa M M, Comeaua J P, et al. Solid-State Electronics, 2006, 50(2): 181

7 Praveen K C, Pushpa N, Cressler J D, et al. Journal of Nano and Electronic Physics, 2011, 3(1): 348

8 Lu W, Yu X, Ren D, et al. Nuclear Techniques, 2005, 28(12): 925 (in Chinese)

9 Schmidt D M, Wu A, Schrimpf R D, et al. IEEE Trans Nucl Sci, 1996, 43(6): 3032

10 Boch J, Saigne F, Touboul A D, et al. Appl Phys Lett, 2006, 88: 232113

11 Witczak S C, Schrimpf R D, Fleetwood D M, et al. IEEE Trans Nucl Sci, 1997, 44(6): 1989

12 Tsetseris L, Schrimpf R D, Fleetwood D M, et al. IEEE Trans Nucl Sci, 2006, 52(6): 2265

13 Hjalmarson H P, Pease R L, Devine R A. IEEE Trans Nucl Sci, 2008, 55(6): 3009

14 Pease R L, Schrimpf R D, Fleetwood D M. IEEE Trans Nucl Sci, 2009, 56(4): 1894

15 Fleetwood Z E, Cardoso A S, Song I, et al. IEEE Trans Nucl Sci, 2014, 61(6): 2915